%% file: main.tex
\documentclass[aps,reprint,twocolumn,superscriptaddress,10pt]{revtex4-1}
\usepackage[dvipdfmx]{graphicx}
\usepackage{siunitx}
\usepackage{hyperref}
\usepackage{amsfonts}
\usepackage{amsmath}
\usepackage{amssymb}
\usepackage{bm}
\usepackage{graphicx}
\usepackage{dcolumn}
\usepackage{euscript}
\usepackage{multirow}
\usepackage{color}
\usepackage{textcomp}
\usepackage{gensymb}
\usepackage{enumitem}
\usepackage{xcolor}
\usepackage{sepfootnotes}

\graphicspath{{./}{../figs/}{./figs/}}

\newcommand{\req}[1]{Eq.~(\ref{#1})}

\newcommand{\rfig}[1]{Fig.~\ref{#1}}
\newcommand{\rFig}[1]{Figure~\ref{#1}}
\newcommand{\rtbl}[1]{Table~\ref{#1}}

\newcommand{\qsgw}{\text{QSGW}}

\newcommand\abinitio{\emph{ab initio}}
\newcommand\etal{\textit{et al.}}

\newcommand{\tj}{\tilde{J}}

\newcommand{\ud}{\,\mathrm{d}}

\newcommand{\cri}{\text{CrI}_3}

\newcommand{\xqwf}{\chi\left({\bf q}, \omega \right)}
\newcommand{\xqwtf}{\chi\left({\bf q}, \omega \right)}

\newcommand{\xqwtb}{\chi_{0}\left( {\bf q}, \omega \right)}
\newcommand{\ispone}{\downarrow}
\newcommand{\isptwo}{\uparrow}
\newcommand{\bfr}{{\bf r}}
\newcommand{\bfq}{{\bf q}}
\newcommand{\bfk}{{\bf k}}

\newcommand{\bfra}{{\bf r}_1}
\newcommand{\bfrb}{{\bf r}_2}

\begin{document}
\title{Electron correlation effects on exchange interactions and spin excitations in 2D van der Waals materials}
\author{Liqin Ke}
\email[Corresponding author: ]{liqinke@ameslab.gov}
\affiliation{Ames Laboratory, U.S.~Department of Energy, Ames, Iowa 50011, USA}

\author{Mikhail I. Katsnelson}
\affiliation{Institute for Molecules and Materials, Radboud University, Heijendaalseweg 135, 6525 AJ Nijmegen, The Netherlands}
\date{\today}



\begin{abstract}
Despite serious effort, the nature of the magnetic interactions and the role of electron-correlation effects in magnetic two-dimensional (2D) van der Waals materials remain elusive.
Using CrI$_3$ as a model system, we show that the calculated electronic structure including nonlocal electron correlations yields spin excitations consistent with inelastic neutron scattering measurements.
Remarkably, this approach identifies an unreported correlation-enhanced interlayer super-superexchange,
which rotates the magnon Dirac lines off, and introduces a gap along, the high-symmetry $\Gamma$-$K$-$M$ path.
This discovery provides a different perspective on the gap opening mechanism observed in CrI$_3$, which was previously associated with spin-orbit coupling through the Dzyaloshinskii-Moriya interaction or Kitaev interaction.
Our observation elucidates the critical role of electron correlations on the spin ordering and spin dynamics in magnetic van der Waals materials and demonstrates the necessity of explicit treatment of electron correlations in the broad family of 2D magnetic materials.
\end{abstract}

\maketitle	

\section{Introduction}
The intimate interplay between correlated electrons, lattice, and magnetism can result in a rich variety of interesting and important physical phenomena~\cite{kotliar2004pt}.
In two-dimensional (2D) materials with open $d$- or $f$-shells, additional quantum confinement caused by the reduced dimensionality suppresses the screening~\cite{wehling2011,MSandMK2011}, and thus may further enhance the electron correlation.
The recent development of magnetic 2D van der Waals (vdW) materials adds a new magnetic functionality to the already vast appeal of the 2D materials family~\cite{gong2017n,huang2017n,burch2018n,gibertini2019nn}.
Their magnetism is very sensitive to, and can be controlled by, pressure~\cite{song2019nm}, stacking arrangement~\cite{klein2018s}, and external magnetic~\cite{gong2017n} and electric~\cite{jiang2018nm,jiang2018nn,deng2018n} fields.
Such tunability offers opportunities to design and construct new energy-efficient spin-based devices.
Even in their bulk form, the reduced coordination number in quasi 2D lattices constrains the electron hopping, thereby increasing the role of the Coulomb interaction.
The resulting enhancement of electron correlations directly affects exchange interactions. 
The critical question is how the magnetic excitations in these confined systems can be understood, controlled, and exploited.

Despite considerable attention, accurate $\abinitio$ descriptions of spin excitations and a comprehensive understanding of magnetic interactions in magnetic 2D vdW materials (m2Dv) are still lacking.
Some complexity is imparted by the role of spin-orbit coupling (SOC).
For example, the presence and role of the Dzyaloshinskii-Moriya interaction (DMI), which is essential for materials to host topological magnons and other topological objects such as skyrmions~\cite{behera2019apl,park2019misc}, remains puzzling.
Specifically, 2D honeycomb ferromagnets can be viewed as the magnetic analog of graphene~\cite{pershoguba2018prx,chen2018prx}.
Earlier theoretical work~\cite{owerre2016jpcm,kim2016prl} demonstrated that introducing a next-nearest-neighboring DMI interaction, which breaks the inversion symmetry of the 2D honeycomb lattice, can induce a spinwave (SW) gap at the Dirac points and allow topological magnons.
Recent inelastic neutron scattering (INS) experiments have shown a gap opening along the high symmetry lines in pristine $\cri$~\cite{chen2018prx}, implying a sizable DMI or Kitaev interaction~\cite{lee2020prl} may indeed exist in $\cri$.
In contrast, $\abinitio$ investigations show that both DMI~\cite{liu2018prb,ghosh2019pbcm,ghosh2020apl,kvashnin2020arxiv} and Kitaev interactions~\cite{xu2018cm,kvashnin2020arxiv} in pristine $\cri$ are insufficient to open up the sizable gap observed in experiments.
The role of the relativistic magnetic interactions is still not fully understood.
However, irrespective of this, the exchange interactions should be reinvestigated beyond density functional theory (DFT) to address the electron-correlation effects~\cite{menichetti20192m,lee2020prb,lee2020xray}.

In this work, using the most studied m2Dv---$\cri$---as a prototype, we identify the role of electron correlations on the magnetic interactions and excitations in m2Dv.
The central quantity that characterizes the spin excitations---the dynamic transverse spin susceptibility (DTSS)---is calculated and directly compared with the SW spectra measured by INS~\cite{chen2018prx,chen2020prb}.
The electron-correlation effects beyond DFT are included via the quasiparticle self-consistent $GW$ (QSGW) method~\cite{van-schilfgaarde2006prl, kotani2007prb}, as incorporated through its effective one-particle potential, which consists of both on-site and off-site non-local parts. 
We demonstrate that the explicit treatment of electron correlations is required to accurately describe the magnetic interactions, especially the interlayer interactions, in m2Dv.
Furthermore, we made the remarkable discovery that a sizable magnon gap opens along the high-symmetry line in $\cri$ even without relativistic exchanges.
Instead, this gap is caused by a correlation-enhanced interlayer Cr-I-I-Cr super-superexchange coupling.

\section{{RESULTS AND DISCUSSION}}
\subsection{Linear response theory}
Starting from a self-consistent $\abinitio$ band structure, we first calculate the bare transverse spin susceptibility $\chi_0(\bfr,\bfr',\bfq,\omega)$ using a linear response method~\cite{kotani2008jpcm,karlsson2000prb,ke2011prbr}.
Then the full transverse susceptibility $\chi$ is calculated, within the random phase approximation (RPA), as $\chi=\chi_0 + \chi_0 I \chi$, where $I$ is the exchange-correlation kernel.
Two-particle quantities $\chi_0$, $\chi$, and $I$ are functions of coordinates $\bfr$ and $\bfr'$ within the unit cell.
Since magnetic moments and excitations are nearly completely confined within the Cr sites, we calculate $\chi_0$ on a product basis~\cite{aryasetiawan1994prbA,kotani2007prb} and then project it onto the local spin densities of Cr pairs.
This projection discretizes $\chi_0(\bfr,\bfr'\bfq,\omega)$ into a matrix $\chi_0(i,j,\bfq,\omega)$, where $i$ and $j$ index the Cr sites in the unit cell.
Such discretization allows us to 1) determine the kernel $I$ using a sum rule~\cite{kotani2008jpcm}; 2) map $\chi^{-1}$ into the Heisenberg model to extract pair exchange parameters; and 3) greatly reduce the computational effort.
As for the electronic structures, we employ the QSGW method~\cite{van-schilfgaarde2006prl, kotani2007prb},
wherein nonlocal potentials, both on-site and long-range off-site, are explicitly calculated.
The widely used DFT$+U$ methods~\cite{AALreview}, which provide a simplistic  correction of on-site Coulomb interactions, are also employed for comparison.
Besides the nonlocal on-site potential, the off-site potential in QSGW could also be critical as it directly affects the relative positions of cation-$3d$ and anion-$p$ bands~\cite{lee2020prb} and thus the indirect exchange interactions of Cr ions via one (superexchange) or multiple (super-superexchange) Iodine sites.
Further details of the QSGW method and implementation~\cite{van-schilfgaarde2006prl, kotani2007prb, pashov2020cpc, kotani2014jpsj} and applications on $\cri$~\cite{lee2020prb} can be found in the Supplementary methods.

By far, except for a few studies~\cite{besbes2019prb,kashin20202dm,kvashnin2020arxiv} using the magnetic force theorem (MFT)~\cite{LKAG1987}, most of the theoretical investigations of the magnetic interactions in m2Dv are based on mapping the total energies of various collinear spin configurations into the Heisenberg model (referred to hereafter as the energy-mapping method), often employed with DFT+$U$.
However, the applicability and accuracy of such an approach in $\cri$ are not clear.
Specifically, using such an energy-mapping method, DFT overestimates the exchange couplings by 50\%~\cite{zhang2015jmcc,chen2018prx}; the additional $U$ correction on Cr sites in DFT+$U$ further increase coupling, only worsening the agreement with experiments.
In principle, the MFT approach is more suitable to describe the small spin deviation from the ground state, such as the SW excitations; moreover, it also allows one to resolve coupling into orbital contributions and elucidate the underlying exchange mechanism~\cite{kashin20202dm,yoon2019jpcm}.
However, the resulting values still vary and are inconclusive; even the opposite trend of exchange-coupling dependence on on-site Cr $U$ values have been obtained~\cite{kvashnin2020arxiv}.
The discrepancy likely lies in the details of the constructions of the TB Hamiltonian and Green’s function~\cite{ke2019prb}, which are often assisted by the Wannier function approach~\cite{mostofi2014cpc}.
On the other hand, DTSS is challenging to compute in practice.
Studies to date have been mostly limited to simple systems, likely due to its computationally-demanding nature and other complications~\cite{okumura2019prb}.
For example, the evaluation of kernel $I$ is not explicit in the case beyond the mean-field scheme, and the Goldstone theorem, which ensures SW at $\bfq=0$ and $\omega=0$ in the absence of magnetocrystalline anisotropy, is often not guaranteed.
In this work, we calculate DTSS without including SOC; the Goldstone magnon mode at $\bfq=0$ and $\omega=0$ is ensured as the kernel $I$ is determined to satisfy the sum rule (Eq.~(2) in ``Supplementary Methods'').
Finally, to investigate the detailed SW dispersion near the Dirac point, high-resolution SW spectra are needed.
This is achieved by calculating the real-space bare susceptibility $\chi_0({\bf R},\omega)$ on a ${\bf R}$-mesh first and then Fourier transforming $\chi_0({\bf R},\omega)$ to obtain $\chi_0(\bfq,\omega)$ with a dense set of $\bfq$ points along high-symmetry paths.

\begin{figure}[bt]
\centering
\begin{tabular}{c}
  \includegraphics[width=1.0\linewidth,clip]{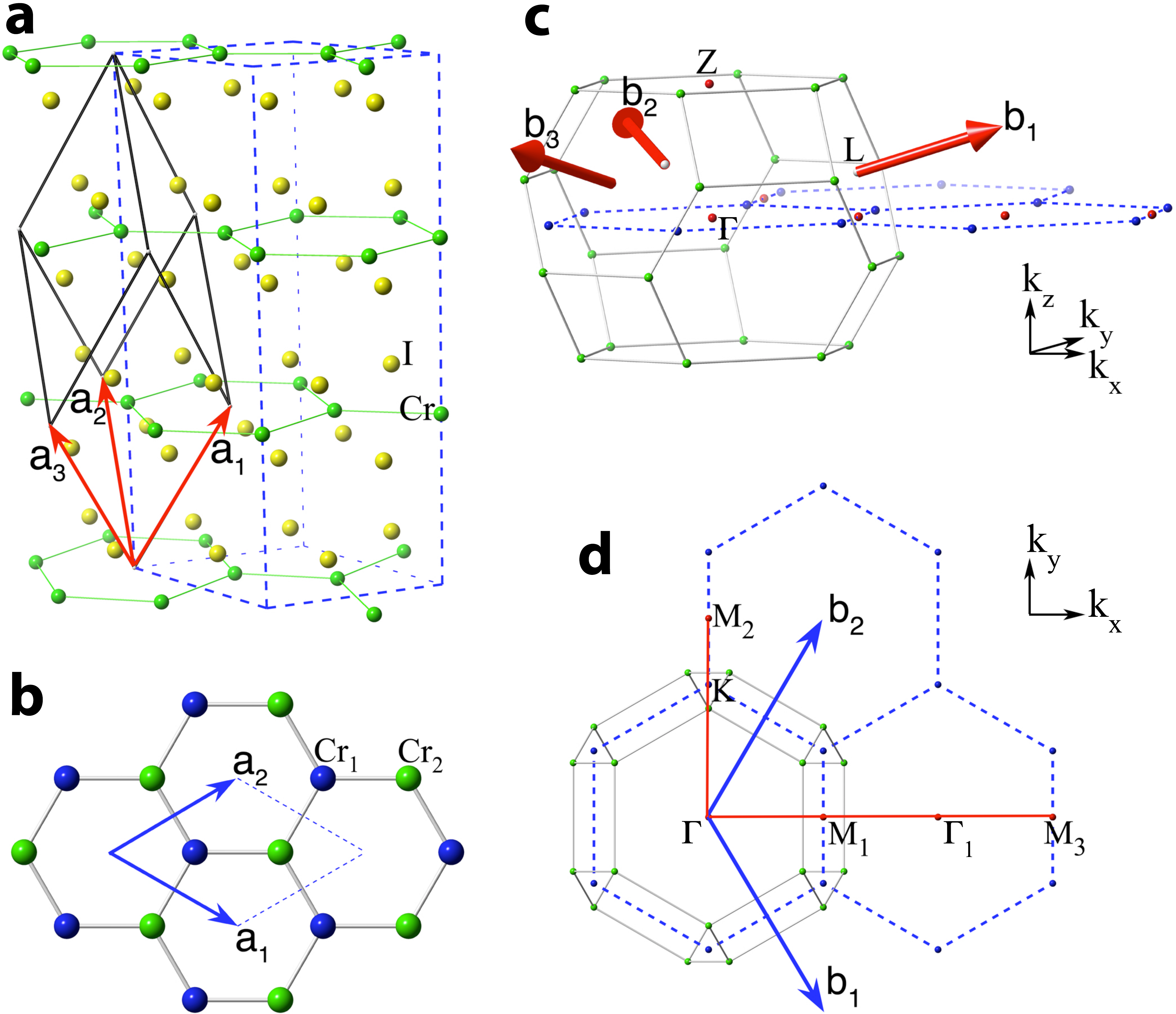}
\end{tabular}%
\caption{\textbf{Crystal structure of  R-$\cri$ and corresponding BZ in comparison to hexagonal lattice.}
\textbf{a} The rhombohedral primitive unit cell and hexagonal conventional unit cell of R-$\cri$ ($R\overline{3}$, BiI$_3$-type, space group no.~148).
\textbf{b} The primitive unit cell of 2D honeycomb Cr sublattice.
\textbf{c} The first BZ and the reciprocal lattice vectors of the rhombohedral primitive cell. 
Along the $\hat{k}_z$ direction, the rhombohedral BZ boundary $Z$ is denoted as (0.5, 0.5, 0.5) in rhombohedral notation or  (0, 0, 1.5) in hexagonal notation.
Dashed lines denote the BZ of the 2D hexagonal lattice.
\textbf{d} The top view of \textbf{c}.
Vectors ${\bf b}_1$ and ${\bf b}_2$ are the reciprocal lattice vectors of the 2D hexagonal lattice shown in \textbf{b}.
}
\label{fig:xtal}
\end{figure}

$\cri$ crystallizes in either the low-temperature rhombohedral structure (R-$\cri$) or high-temperature monoclinic structure (M-$\cri$)~\cite{mcguire2015cm}.
A honeycomb Cr monolayer is sandwiched between two I layers; then, the blocks of I-Cr-I triple layers are stacked along the $z$ direction, held together by a weak vdW force. 
M-$\cri$ has a slightly distorted honeycomb Cr lattice and, more importantly, a different stacking arrangement, resulting in the $A$-type antiferromagnetic (AFM) ordering~\cite{klein2018s,sivadas2018nl,soriano2019ssc,soriano2020prb} instead of the ferromagnetic (FM) ordering as in R-$\cri$.
The sensitivity of interlayer Cr coupling on stacking arrangement reflects the changes of super-superexchange pathways across the vdW gap.
A thorough understanding requires an explicit treatment of interlayer exchanges, instead of using a single effective interlayer exchange parameter as is often employed to describe the system.
Thus, the exchange couplings and SWs in R-$\cri$ need to be considered in the context of rhombohedral symmetry.
In this work, we first focus on the magnetism in FM R-$\cri$ and then discuss stacking effects on magnetism using M-$\cri$.

\rFig{fig:xtal} shows the crystal structure of R-$\cri$ and the corresponding Brillouin zone (BZ).
The 2D honeycomb Cr sublattice and corresponding BZ are also shown for better comparison with previous studies, in which the SWs are often discussed in the hexagonal notation.
The rhombohedral primitive unit cell contains two formula units (f.u.), while the conventional hexagonal cell includes six.
Correspondingly, the former has a larger BZ than the latter.
As shown in \rfig{fig:xtal}(d), along the $k_x$ direction, its BZ boundary $M_3$ is three times further from $\Gamma$ than that of the hexagonal structure, $\Gamma\textendash M_1$.
Along the $k_y$ direction, $M_2$ (equivalent to $M_3$) is at the BZ boundary for both rhombohedral and hexagonal cells.

\begin{figure*}[thb]
\centering
\includegraphics[width=1.0\linewidth,clip]{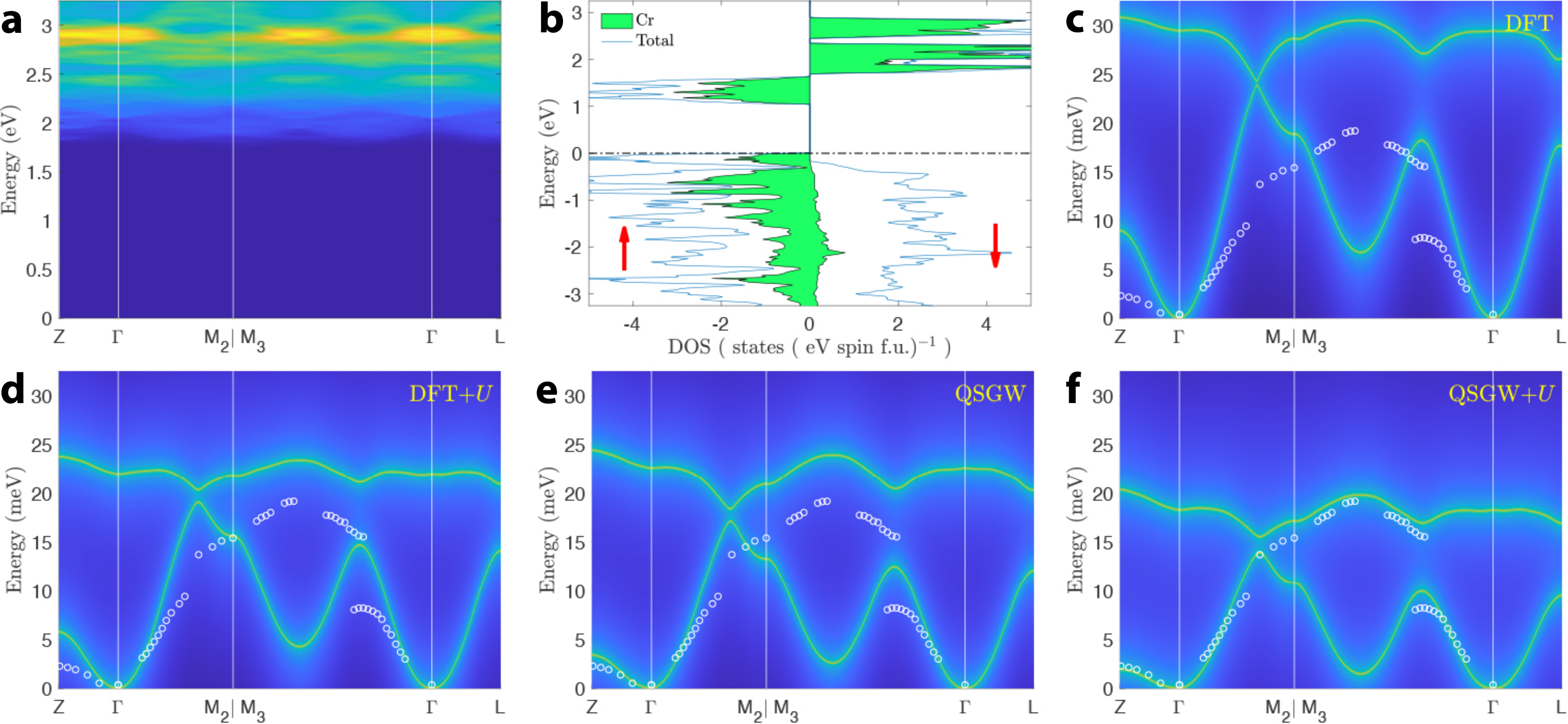}   
\caption{\textbf{Spin excitations calculated from $\operatorname{Im}\bm{[\xqwtb]}$ and  $\operatorname{Im}[\bm{\xqwtf]}$ in R-$\cri$.}
  The special $\bfq$ points along the high symmetry path $Z$--$\Gamma$--$M_2|M_3$--$\Gamma$--$L$ are denoted in Figs.~\ref{fig:xtal}\textbf{c} and \ref{fig:xtal}\textbf{d}.
\textbf{a} $\operatorname{Im}[\xqwtb]$ calculated in DFT.
\textbf{b} The density of states calculated in DFT. The horizontal dashed-and-dotted line indicates the top of valence band.
\textbf{c}--\textbf{f} $\operatorname{Im}[\xqwtf]$ calculated in DFT (\textbf{c}), DFT$+U$ (\textbf{d}), QSGW (\textbf{e}), and QSGW$+U$ (\textbf{f}). The intensity of $\operatorname{Im}[\xqwtf]$ is shown in log scale.
Experimental SW energies, adopted from INS work by chen $\etal$~\cite{chen2018prx,chen2020prb}, are denoted by open circles.  
}
\label{fig:x0x2}
\end{figure*}

\subsection{Dynamic transverse spin susceptibility}

First, we calculate within DFT the bare and full transverse susceptibilities, $\xqwtb$ and $\xqwtf$, which characterize the single-particle Stoner excitations and collective SW excitations, respectively.
The intensities of $\operatorname{Im}[\xqwtb]$ and $\operatorname{Im}[\xqwtf]$ along high-symmetry paths are shown in \rfig{fig:x0x2}(a) and \ref{fig:x0x2}(c), respectively.
The corresponding special ${\bf q}$ points are denoted in \rfig{fig:xtal}(c) and \ref{fig:x0x2}(d).
The energy scale of the Stoner excitations, starting at $\sim$\SI{1.7}{eV} and peaking at $\sim$\SI{3}{eV}, are two orders of magnitude larger than SW excitations, leaving the latter with no damping.
This means that we can likely use the physical picture of well-defined local magnetic moments on Cr atoms and map the low-lying spin dynamics onto a purely localized-spin Hamiltonian as will be described in detail below.
The threshold energy of $\sim$\SI{1.7}{eV} corresponds to the gap size of the spin-flip transition from the top of majority-spin Cr valence states to the bottom of minority-spin Cr conduction states, as shown in \rfig{fig:x0x2}(b), while the peak energy corresponds to the spin splitting of the Cr-$d$ states.
The SW energies, defined by the peaks of $\operatorname{Im}[\xqwtf]$, are solely determined by the RPA poles of $1-I\chi_0=0$.
With two Cr atoms in the primitive cell, we find two poles for each $\bfq$, resulting in two magnon branches, as shown in Figs.~\ref{fig:x0x2}(c)--\ref{fig:x0x2}(f).
Experimental SW energies extracted from previous INS work by Chen $\etal$~\cite{chen2018prx,chen2020prb} are plotted to compare with the calculated SW spectra.

We now compare the DFT SW spectra with INS experiments.
Along the $\Gamma$--$M_2$ path, \rfig{fig:x0x2}(c) shows that two SW branches cross near the point $K$, before reaching the BZ boundary $M_2$.
For other directions, a gap exists between two magnon branches.
This is consistent with previous studies~\cite{chen2018prx} without considering DMI.
Along the $\Gamma$--$M_3$ path, the SW minimum occurs at $[3\,0\,0]$ (hexagonal notation), instead of $[1\,0\,0]$ ($\Gamma_1$-point), reflecting the symmetry of the rhombohedral structure.
The maximum energy of the optical mode measured in INS is about \SI{20}{\meV}, while DFT gives \SI{30}{\meV} and overestimates it by 50\%.
As shown in \rfig{fig:x0x2}(c),  DFT also overestimates the acoustic magnon energies, most severely (by a factor of $\sim$4) along the confined $z$ direction.
The overestimation of interlayer coupling also affects the in-plane SW energies, because nearly all interlayer couplings have in-plane components in their connecting vectors.
Hence, in bulk materials, an accurate description of interlayer exchanges is also essential to describe the in-plane SW accurately.

Next, we investigate the effects of electronic correlations on magnetic interactions within DFT$+U$ and QSGW.
We found that increasing the $U$ value in DFT$+U$ increases the on-site Cr moment  (Supplementary Table I) and lowers the energies of both magnon branches (Supplementary Fig.~3).
\rFig{fig:x0x2}(d) shows the SW spectra for $U=\SI{3}{\eV}$, a typical value used for $\cri$.
The SW energy at the $Z$ point is decreased to \SI{5.8}{meV}, about 2/3 of the DFT value, however, it is still about a factor of three larger than that found in INS experiments.
Moreover, the optical mode becomes flatter, in contrast to INS~\cite{chen2018prx}.
Surprisingly, electron-correlation effects also introduce a SW gap at the Dirac point, which will be discussed in detail later.
\rFig{fig:x0x2}(e) shows the SW spectra calculated within QSGW.
The optical magnon is centered at around \SI{24}{\meV}, similar as in DFT$+U$, but recovers some dispersion and agrees better with INS experiments.
Interestingly, the acoustic SW energies are reduced in comparison to DFT$+U$.
The SW energy at the $Z$-point drops to $\sim$\SI{3.4}{meV}, showing a much weaker interlayer FM coupling and better agreement with experiments.
Thus, the elaborate treatment of electron interactions in QSGW improves the description of spin excitations in these systems.
Considering $GW$ methods may underestimate the on-site electron interactions, they have been applied on top of DFT+$U$ for various systems~\cite{jiang2012prb}.
Here, we also apply QSGW on top of DFT+$U$ (referred to hereafter as QSGW+$U$)~\cite{sponza2017prb} to roughly mimic the additional on-site interactions.
\rFig{fig:x0x2}(f) shows the SW spectra calculated with $U=\SI{1.36}{\eV}$.
Although the in-plane acoustic SW is still somewhat overestimated, the overall spectra compare well with experiments, 
suggesting that additional on-site interactions beyond QSGW may be needed to best describe electronic structures and SW in $\cri$. 
More rigorous and comprehensive frameworks, such as the dynamical mean-field theory (DFMT)+$GW$~\cite{tomczak2012prl}, could prove valuable for future research.

\subsection{Exchange parameters}

\begin{table}[htb]
  \caption{\textbf{Pairwise intralayer ($\bm{J_{i}}$) and interlayer ($\bm{\tilde{J}_{i}}$) exchange parameters in R-$\cri$ calculated within various methods.}
The Heisenberg model is defined as $H=-\sum_{i\ne j} J_{ij}\,\hat{\bf e}_i \cdot \hat{\bf e}_j$, where $\hat{\bf e}_i$ is the unit vector of the atomic spin moment at site $i$.    
The degeneracy (No.) and distance ($R_{ij}$) of $J_{ij}$ are also provided.
Symbol * denotes intra-sublattice couplings, which do not contribute the magnon gap between acoustic and optical modes.
Positive (negative) $J_{ij}$ values correspond to FM (AFM) couplings.
$U=\SI{3}{\eV}$ and $U=$\SI{1.36}{\eV} are used in DFT+$U$ with QSGW+$U$ calculations, respectively.
$\tilde{J}_{2_0}$ vanishes in all calculations.
  }
\label{tbl:jij}%
\bgroup
\def\arraystretch{1.05}%
\small
\begin{tabular*}{\linewidth}{llc@{\extracolsep{\fill}}ccccccc}
  \hline  \hline
  \\[-1em]
  \multicolumn{2}{c}{R-$\cri$} & & $R_{ij}$  &   & \multicolumn{4}{c}{$J_{ij}$(meV)}\\
  \\[-1.1em]
  \cline{1-2} \cline{4-4}  \cline{6-9}
  \\[-1.1em]
  Lbl.  & No. & & \AA & & DFT & DFT+$U$ & QSGW & QSGW+$U$ \\
  \\[-1.1em]
  \hline
  \\[-1.1em]
      $J_{1}$           & 3  & &  3.965   &  & 3.29  &  2.67 &  2.84 &  2.48  \\
      $J_2$             & 6* & &  6.867   &  &  0.57 &  0.61 &  0.43 &  0.40  \\
      ${J}_{3}$         & 3  & &  7.929   &  & -0.07 &  0.02 & -0.06 & -0.07  \\
  \\[-1.1em]  
  \hline
  \\[-1.1em]
      $\tilde{J}_{0}$   & 1  & &  6.589   &  &  0.19 &  0.25 &  0.14 &  0.14   \\  
      $\tilde{J}_{1-}$  & 6* & &  7.701   &  &  0.31 &  0.25 &  0.16 &  0.13       \\
      $\tilde{J}_{1+}$  & 3  & &  7.713   &  &  0.47 &  0.42 &  0.26 &  0.23       \\
      $\tilde{J}_{2_0}$ & 3  & &  9.517   &  &  0.00 &  0.00 &  0.00 &  0.00       \\
      $\tilde{J}_{2}$   & 3  & &  9.517   &  & -0.04 & -0.19 & -0.18 & -0.25       \\    
  \hline\hline
\end{tabular*}
\egroup
\end{table}

\begin{figure}
\centering
\begin{tabular}{c}
\includegraphics[width=1.0\linewidth,clip]{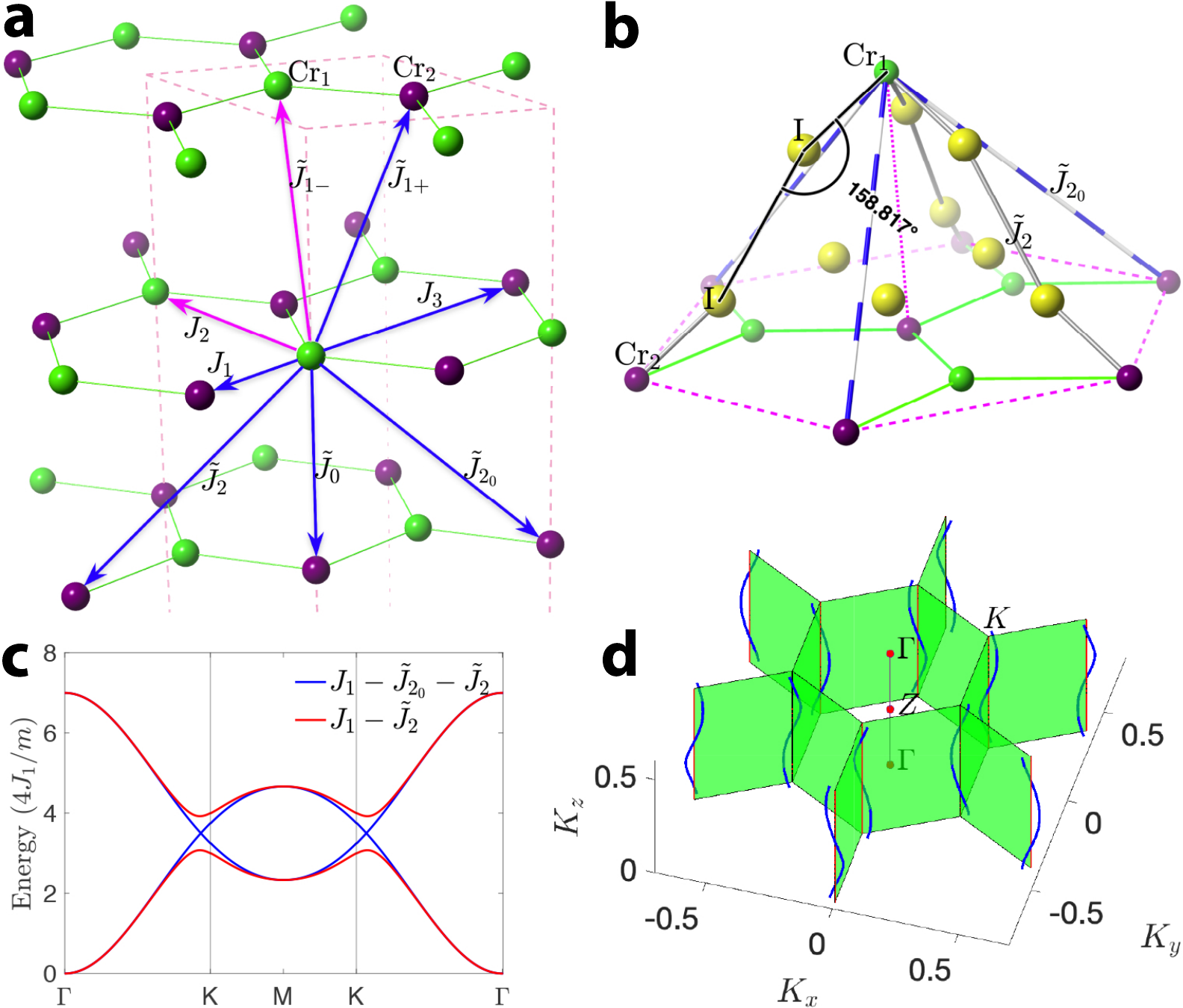}   
\end{tabular}%
\caption{\textbf{Exchange parameters, SW spectra and nodal lines calculated using linear spinwave theory.}
  \textbf{a} Pair exchange parameters for the first few neighbors in R-$\cri$.
    \textbf{b} Atomic configurations around the exchange paths of $\tilde{J}_{2_0}$ and $\tilde{J}_2$.
  AFM $\tilde{J}_{2}$ corresponds to a Cr-I-I-Cr super-superexchange coupling.
  \textbf{c} SW spectra along the $\Gamma$--$K$--$M$ path calculated within a $J_1$-$\tilde{J}_{2}$-$\tilde{J}_{2_0}$ ($\tilde{J}_{2}=\tilde{J}_{2_0}={J}_{1}/12$) model and a  $J_1$-$\tilde{J}_{2}$ ($\tilde{J}_{2}={J}_{1}/6$)  model.
  \textbf{d} Dirac nodal lines, where the magnon bands cross, wind around the $K$-point and along the $\hat{k}_{z}$ direction; the exchange parameters calculated in QSGW$+U$ are used.
}
\label{fig:Jgap}
\end{figure}

To develop a quantitative understanding of how electron correlations affect magnetic interactions and excitations, we calculate the effective pair exchange parameters $J_{ij}$ for a Heisenberg model $H=-\sum_{i\ne j} J_{ij}\,\hat{\bf e}_i \cdot \hat{\bf e}_j$, where $\hat{\bf e}_i$ is the unit vector of the atomic spin moment at site $i$.
The $J_{ij}$ parameters are obtained from the inverse of susceptibility matrix, $[\chi(\bfq,\omega=0)]^{-1}$, with a subsequent Fourier transform~\cite{szczech1995prl,katsnelson2004jpcm,kotani2008jpcm,ke2013prb}.
As shown in Ref.~\cite{katsnelson2004jpcm}, the exchange parameters determined in this way coincide with those calculated via the magnetic force theorem.
The exchange parameters between the first few neighbors, as depicted in \rfig{fig:Jgap}(a), are listed in \rtbl{tbl:jij}; couplings beyond \SI{12}{\AA} are negligible.
Using the linear SW theory, we recalculate the SW spectra with the extracted $J_{ij}$ parameters.
The obtained spectra agree well with those determined by the peaks of $\operatorname{Im}[\xqwf]$ (See Supplementary Fig.~2).

We find that the on-site Hubbard $U$ corrections on Cr-$d$ states included in DFT$+U$ have a stronger effect on decreasing the nearest-neighboring coupling, while the explicit non-local potentials in QSGW have a more substantial effect on the longer-range interlayer couplings.
Within DFT, the calculated in-plane exchanges are similar to values~\cite{zhang2015jmcc} derived from total energy mapping and $\sim50\%$ larger than those extracted from INS~\cite{chen2018prx}.
In comparison to DFT, DFT$+U$ decreases $J_1$ by $\sim19\%$ and the overall interlayer (FM) couplings $\sum\tilde{J}$ by $\sim27\%$, while QSGW decreases $J_1$ by $\sim14\%$ and $\sum\tilde{J}$ by $\sim60\%$.
QSGW not only reduces the optical SW energies but also significantly lowers the acoustic ones, especially along the interlayer direction.

Interestingly, regarding the dependence of exchange and SW energy on the $U$ parameter,  the energy-mapping method using multiple collinear states gives the opposite trend from the linear response method.
Using the energy-mapping method, exchange and SW energy increase with $U$, worsening the agreement between theory and INS measurements.
This suggests that non-Heisenberg interactions, such as biquadratic exchange or multi-site exchange interactions~\cite{turzhevskii1990spss,logemann2017jpcm,wysocki2011np,wysocki2013jpcs,kartsev2020cm},
might be important in this system, especially when additional electron correlations are taken into account.
By analogy with transition-metal oxide systems~\cite{logemann2017jpcm} one can assume that the induced magnetic polarization on iodine ligands can play a decisive role in this non-Heisenberg behavior. Contrary to the total energy differences, the linear response method describes accurately the small spin deviations from the given (ground) state and thus the spinwave spectra.
The corresponding exchange integrals thus depend on the initial equilibrium spin configuration in which they are calculated~\cite{LKAG1987,turzhevskii1990spss}.
On the other hand, mapping the total energy of spin-spiral configurations may provide a better description of $J_{ij}$ than the collinear configurations.

\subsection{Interlayer-coupling induced gap opening along ${\Gamma}$--${K}$--${M}$}
Remarkably, besides the improvement of SW energies, correlations beyond DFT also open up a gap along the $\Gamma$--$K$--$M_2$ path in both DFT$+U$ and QSGW, as shown in Figs.~\ref{fig:x0x2}(d)--\ref{fig:x0x2}(f).
Although the calculated gap size ($\sim\SI{1.8}{\meV}$ in QSGW+$U$) is smaller than the experimental value of $\sim\SI{4}{\meV}$ observed in INS, the existence of such a gap along the $\Gamma$--$K$--$M$ path is unexpected in the absence of DMI or Kitaev interaction.
As we show later, this gap can be caused by the correlation-enhanced interlayer super-superexchange $\tilde{J}_2$.
If one only considers the Cr sublattice itself, which has a higher symmetry ($R\overline{3}m$), $\tilde{J}_{2_0}$ and $\tilde{J}_2$ should be equivalent.
The presence of the Iodine sublattice breaks the mirror and 2-fold rotational symmetry, lifting the degeneracy of $\tilde{J}_{2_0}$ and $\tilde{J}_2$.
Thus, although $\tilde{J}_{2_0}$ and $\tilde{J}_2$ connect similar Cr pairs with the same distance of \SI{9.517}{\AA}, their exchange paths are not the same, which allows for $J_2$ and $J_{2_0}$ to adopt different values.
However, up until now, all previous work has considered that $J_{2}=J_{2_0}=0$, which indeed is supported, to a great extent, by DFT.
Surprisingly, the inclusion of correlations beyond DFT results in a sizable AFM $\tilde{J}_2$.
As shown in ~\rtbl{tbl:jij}, $\tilde{J}_{2_0}$ vanishes in all calculations, while AFM $\tilde{J}_{2}$ is negligible in DFT but becomes stronger in QSGW and DFT$+U$, reaching $\tilde{J}_2=10\%J_1$ in QSGW$+U$.
As shown in \rfig{fig:Jgap}(b), $\tilde{J}_{2}$  corresponds to a Cr-I-I-Cr super-superexchange with a Cr-I-I angle of $159\degree$, giving the major AFM contribution to interlayer couplings, whereas vanishing $J_{2_0}$ has no obvious exchange path with intervening I anions.

The gap between acoustic and optical modes, at an arbitrary $\bfq$ point, gives the energy difference between the in-phase and out-of-phase precessions of two Cr-spin sublattices and depends on the inter-sublattice couplings $2|B(\bfq)|$ (see \req{eq:sw} and \req{eq:aqbq} of the ``Methods'' section).
Within the considered exchange range, we have $B(\bfq)=J_1(\bfq) + J_3(\bfq) + \tj_{0}(\bfq) + \tj_{1+}(\bfq) + \tj_{2_0}(\bfq) + \tj_{2}(\bfq)$, where $J_i(\bfq)$ is the corresponding Fourier component of $J_i({\bf R})$.
Couplings $J_1(\bfq)$, $J_3(\bfq)$, and $\tj_{1+}(\bfq)$ are real functions along the $\Gamma$--$K$--$M$ path and vanish at the $K$-point, resulting a bandcrossing at $K$ if other terms are ignored.
The interlayer coupling $\tj_{0}$ is along the $z$ direction; $\tj_{0}(\bfq)=\tj_{0}$ is a real constant when $\bfq$ is in the basal plane, shifting the bandcrossing along the $\Gamma$--$K$--$M$ path.
In real space, the connecting vectors (in-plane components) of $\tj_{2_0}$ and $\tj_{2}$ are rotated by $\pi/6$ with respect to those of  $J_1$, $J_3$, and $\tj_{1+}$.
Correspondingly, in reciprocal space, $\tj_{2_0}(\bfq)$ and $\tj_{2}(\bfq)$  are complex functions along the $\Gamma$--$K$--$M$ path  (See Supplementary Fig.~4).
With $\tj_{2_0}=0$, $\tj_{2}(\bfq)$ itself results in a non-vanishing $|B(\bfq)|$ and thus a gap along the $\Gamma$--$K$--$M$ path.
However, if $\tj_{2_0}=\tj_{2}$, then $(\tj_{2_0}(\bfq)+\tj_{2}(\bfq))$ is a real function when $\bfq$ is in the basal plane, shifting the magnon crossing as $\tj_{0}$ does along the $\Gamma$--$K$--$M$ path. Thus, the combination of vanishing $\tj_{2_0}$ and correlation-enhanced $\tj_{2}$ will induce the magnon gap along the $\Gamma$--$K$--$M$ path.
To illustrate, we calculate the SW spectra in a simple $J_1$--$\tilde{J}_{2_0}$--$\tilde{J}_2$ model with $\tilde{J}_{2_0}=\tilde{J}_{2}=J_1/12$ and a $J_1$--$\tilde{J}_2$ model with $\tilde{J}_{2}=J_{1}/6$, respectively.
As shown in \rfig{fig:Jgap}(c), the latter case opens a gap along the $\Gamma$--$K$--$M$ path.

However, unlike the global DMI-induced gap, the gap induced by the nonequivalence of the exchange interactions $\tilde{J}_{2_0}$ and $\tilde{J}_{2}$ does not persist through the whole BZ because a solution of $B(\bfq)=0$ can be found near the $K$-point.
The Dirac nodal lines in the SW spectra still form but do not cross the $\Gamma$--$K$--$M$ lines (at the $k_z=0$ plane).
Using the $J_{ij}$ parameters obtained within QSGW+$U$, we calculate,  within the linear SW theory, the in-plane SW spectra at various $k_z$ planes.
\rFig{fig:Jgap}(d) shows the helical Dirac nodal lines form around the edges of the hexagonal BZ; each line crosses only twice the face of the first BZ.
It would be interesting to see whether future INS experiments can confirm the small displacement of the Dirac point off the $\Gamma$--$K$ line or at finite $k_z$.
However, such measurement may be challenging due to the complexity from the modulation of the dynamic structure factor close to the gap, the requirements of high instrumental resolution and good sample mosaic, and the finite SOC effects.

\subsection{Stacking-dependent magnetic ordering}
\begin{figure}
\centering
\includegraphics[width=1.0\linewidth,clip]{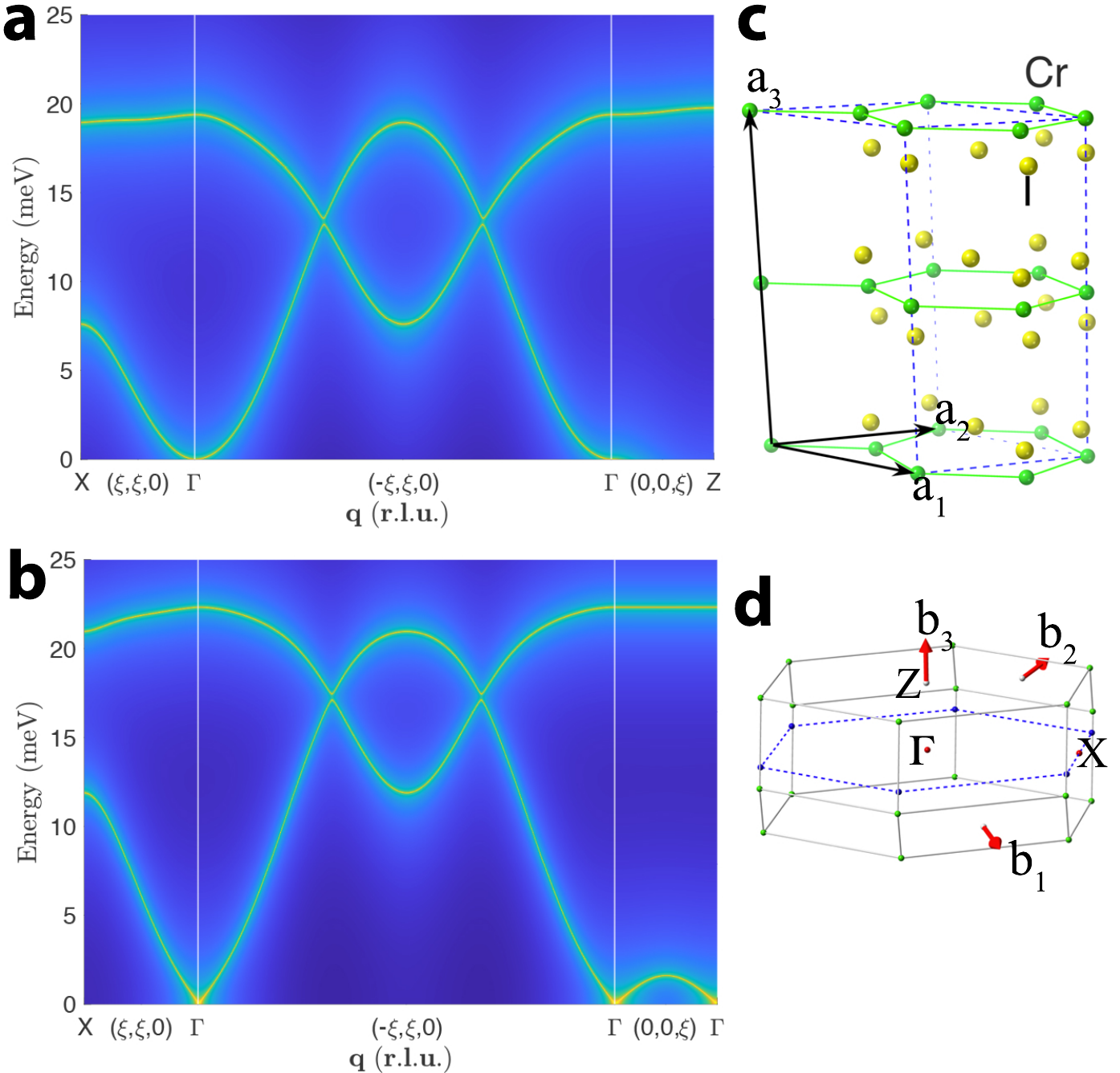} 
\caption{\textbf{SW spectra in FM and $\bm{A}$-type AFM M-$\cri$ calculated within QSGW.}
  Wavevectors $\bfq=h{\bf b}_1 + k{\bf b}_2 + l{\bf b}_3$ is denoted as $\bfq=(h, k, l)$ in reciprocal lattice units (r.l.u.).
  SW are plotted along $X$--$\Gamma$--$Y$--$\Gamma$--$Z$ for the FM structure and $X$--$\Gamma$--$Y$--$\Gamma$--$Z$--$\Gamma$ for the AFM structure.
  High-symmetry $\bfq$ points $X$, $Y$, and $Z$ are at BZ boundaries and denoted as (0.5, 0.5, 0), (-0.5, 0.5, 0), and (0, 0, 0.5), respectively.
\textbf{a} $\operatorname{Im}[\xqwtf]$ calculated in FM configuration. 
\textbf{b} $\operatorname{Im}[\xqwtf]$ calculated in AFM configuration. 
\textbf{c} The primitive cell of AFM M-$\cri$ structure, in which the lattice vector ${\bf a}_3$ is doubled, in comparison to the FM structure.
\textbf{d} The first BZ of M-$\cri$. Reciprocal lattice vectors ${\bf b}_1$ and  ${\bf b}_2$ are the same for both FM and AFM structures while ${\bf b}_3^{(\text{FM})}=2{\bf b}_3^{(\text{AFM})}$.
}
\label{fig:x2_mono}
\end{figure}
Finally, we demonstrate that explicit treatments of electron correlations can correctly describe the dependence of interlayer interaction on stacking order.
M-$\cri$ has different stacking than R-$\cri$, which dramatically modifies the inter-layer super-superexchange paths and results in $A$-type AFM ordering in M-$\cri$.
This intimate interplay between stacking order and magnetic ordering plays a crucial role in manipulating the magnetism in these materials.
DFT total energy calculations predict the wrong FM ground state for M-$\cri$, while DFT+$U$ calculations~\cite{sivadas2018nl,soriano2019ssc} have shown that AFM interlayer configurations can be stabilized in M-$\cri$, depending on the $U$ value.
Within QSGW, we calculate $\xqwtf$ in M-$\cri$ starting from both the FM and the $A$-type AFM ground-state configurations.
The corresponding SW spectra along the high symmetry paths are shown in \rfig{fig:x2_mono}(a) and \ref{fig:x2_mono}(b), respectively.
The acoustic SW calculated with FM configuration, as shown in \rfig{fig:x2_mono}(a), is negative along the $\Gamma$--$Z$ path (normal to the basal plane), suggesting the instability of the FM interlayer configuration in M-$\cri$.
In contrast, the SW spectra calculated with the AFM configuration, as shown in  \rfig{fig:x2_mono}(b), are positive through the whole BZ.
Thus, by taking into account the explicit electron correlations in QSGW, the magnetic ordering dependence on stacking can be correctly described in a parameter-free fashion.

In conclusion, we have demonstrated that an $\abinitio$ description of electron correlations beyond DFT can accurately describe the SW spectra and the stacking-dependent magnetism in a parameter-free fashion, and reveals the gap opening mechanism in $\cri$.
An unexpected correlation-enhanced interlayer super-superexchange induces a gap along the $\Gamma$--$K$--$M$ path in the absence of previously proposed relativistic exchanges or magnon-phonon interactions~\cite{kvashnin2020arxiv}.
To experimentally verify the true physical mechanism of the gap opening and the nature of magnetic interactions, future INS experiments may be used to search for the bandcrossings off the high-symmetry line in bulk $\cri$.
Identifying the existence of the magnon gap along the $\Gamma$--$K$--$M$ path in monolayer $\cri$, in which the interlayer exchange is absent, can also help illuminate the responsible interactions.
Of course, INS studies of single-layer materials seem to be impossible due to a small number of atoms.
However, other techniques such as electron energy loss spectroscopy can be, in principle, used~\cite{senga2019}.
Our work suggests the necessity of an explicit treatment of electron correlations to accurately describe the magnetism in the broad family of layered magnetic materials~\cite{cenker2020np,zhang2020misc}, including magnetic topological vdW materials~\cite{li2020prl,li2020twodimensional}, where the interaction between magnetization and the topological surface state is essential.

\section{METHODS}

\subsection{Pair exchange parameters for Heisenberg model}
In adiabatic approximation we map spin susceptibility into a classical Heisenberg model
\begin{equation}
  H=-\sum_{i\ne j} J_{ij}\,\hat{\bf e}_i \cdot \hat{\bf e}_j,
\label{eq:Hamiltonian}  
\end{equation}
where $\hat{\bf e}_i$ is the unit vector of magnetic moment on site $i$.
Effective pair exchange parameters are calculated as
\begin{equation}
  J_{ij} = \frac{1}{4}\int_{\Omega_i}  \ud{\bfr_1} \int_{\Omega_j} \ud{\bfr_2}\, m_i({\bfr_1}) J({\bfr_1},{\bfr_2}) m_j(\bfr_2),
\end{equation}
where $m_i({\bfr})$ is the density of magnetic moment on Cr site $i$ and  $J(\bfr_1,\bfr_2)$ is related to  $\chi^{+-}(\bfr_1,\bfr_2)$  and satisfies
\begin{equation}
\int_{\Omega} \ud{\bfr_2}\, J(\bfr_1,\bfr_2)  \chi^{+-}(\bfr_2,\bfr_3,\omega=0)  = \delta(\bfr_1-\bfr_3).
\end{equation}
Thus, the effective pair exchange parameters can be obtained from the matrix elements of the inverse of spin susceptibilities matrix~\cite{szczech1995prl,katsnelson2004jpcm,kotani2008jpcm} with a subsequent Fourier transform,
\begin{eqnarray}
  &&J_{ij}({\bf R})=  \frac{1}{\Omega _{\rm BZ}}  \int \ud\mathbf{q}\,  e^{i2\pi\mathbf{ q\cdot R}}  J_{ij}(\mathbf{q}) \\ 
  &=&\frac{1}{\Omega _{\rm BZ}}  \int \ud\mathbf{q}\,  e^{i2\pi\mathbf{ q\cdot R}} 
   ||m_i|| \left[ \left( \chi^{+-}(\mathbf{q},\omega=0)\right)^{-1}\right]_{ij} ||m_j||. \nonumber
\end{eqnarray}
Here, the $\chi_{ij}^{+-}$ matrices are obtained by projecting onto the functions $\{m_i(\bfr)/||m_i||\}$ representing normalized local spin densities on each magnetic Cr site, which gives a matrix $\chi _{ij}( \mathbf{q},\omega )$ in  magnetic basis site indices~\cite{kotani2008jpcm}.
This projection corresponds to the rigid spin approximation (RSA), which is a modest approximation for $\cri$.
A $8\times 8 \times 8$ $\bfq$ mesh was employed to calculate $J_{ij}({\bf R})$.
Using a larger $12 \times12 \times 12$ $\bfq$ mesh barely changes the SW spectra (See Supplementary Fig.~1).

The detailed calculation of transverse spin susceptibility using the linear response method and preparation of the single-particle Hamiltonian, including both DFT and QSGW, can be found in ``Supplementary Methods.''

\subsection{Linear spinwave theory }
Starting from the Heisenberg Hamiltonian, \req{eq:Hamiltonian}, and using the Holstein-Primakoff transformation, 
the SW energies of the acoustic and optical modes are written as:
\begin{equation}
  E(\bfq)= \frac{4}{m} \left( A(0)+B(0) -A(\bfq) \pm |B(\bfq)| \right),
  \label{eq:sw}
\end{equation}
where $A(\bfq)$ and $B(\bfq)$ are sum of $J(\bfq)$ over the intra-sublattice and inter-sublattice pairs, respectively:
\begin{equation}
  \begin{array}{rcl}
  A(\bfq)&=&J_2(\bfq)+\tj_{1-}(\bfq)  \\
  B(\bfq)&=&  J_1(\bfq) + J_3(\bfq) + \tj_{0}(\bfq) + \tj_{1+}(\bfq) + \tj_{2_0}(\bfq) + \tj_{2}(\bfq). 
  \end{array}  
  \label{eq:aqbq}  
\end{equation}
Here, 
\begin{equation}
  J_i(\bfq)= J_i \sum_{\delta} e^{-i 2 \pi \bfq\cdot \delta}
\end{equation}
is the Fourier transform of real space exchange parameters $J_i$ with corresponding connecting vectors $\delta$. 
As shown in Eq.~(\ref{eq:sw}), the gap between two modes is $\propto |2B(\bfq)|$.
Along the  $\Gamma$--$K$--$M$ path, $J_1(\bfq)$, $J_3(\bfq)$, and $\tj_{1+}(\bfq)$ are real cosine functions of $\bfq$ and vanish at point $K$, while 
$\tj_0(\bfq)$ is a real constant and $\tj_{2}(\bfq)$ a complex number.

\acknowledgments
 We are indebted to B.~Harmon, R.~J.~McQueeney, A.~Kressig, B.~Li, J.~R.~Morris, and M.~van Schilfgaarde for fruitful discussions.
  We are also grateful to L.~Chen and P.~Dai for providing their INS data.
  This work was supported by the U.S.~Department of Energy, Office of Science, Office of Basic Energy Sciences, Materials Sciences and Engineering Division, and Early Career Research Program. 
  Ames Laboratory is operated for the U.S.~Department of Energy by Iowa State University under Contract No.~DE-AC02-07CH11358.
  This research used resources of the National Energy Research Scientific Computing Center (NERSC), a U.S.~Department of Energy Office of Science User Facility operated under Contract No.~DE-AC02-05CH11231.
  The work by MIK is supported by European Research Council via Synergy Grant 854843 - FASTCORR.  

  

\bibliography{aaa}

\widetext
\clearpage
\textbf{\Large Supplementary information for:}
\newline
\begin{center}
\textbf{\large Electron correlation effects on exchange interactions and spin excitations in 2D van der Waals materials}
\end{center}
\setcounter{equation}{0}
\setcounter{figure}{0}
\setcounter{table}{0}
\setcounter{section}{0}
\makeatletter
\renewcommand{\theequation}{S\arabic{equation}}
\renewcommand{\thefigure}{S\arabic{figure}}
\renewcommand{\thetable}{S\arabic{table}}
\renewcommand{\bibnumfmt}[1]{[S#1]}
\renewcommand{\citenumfont}[1]{S#1}

\vspace{20pt}

\section{Supplementary Figures}

\begin{figure}[h]
\begin{tabular}{c}
\includegraphics[width=0.75\linewidth,clip]{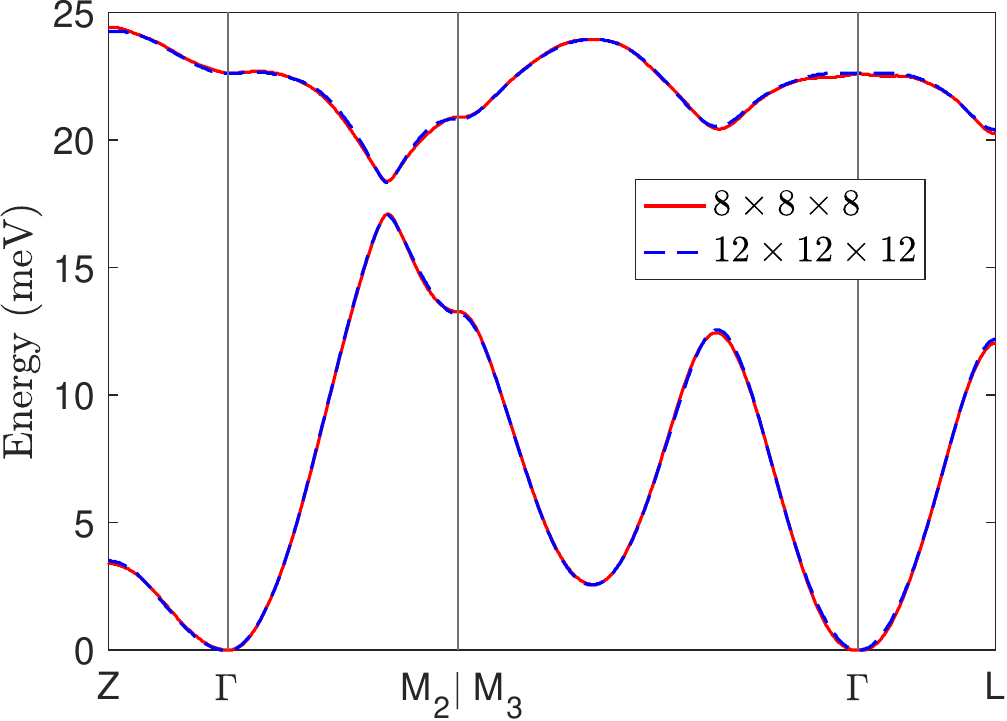} 
\end{tabular}%
\caption{Convergence of ${\chi^{+-} \left( {\bf q}, \omega \right)}$ with respect to $\bm{k}$ mesh.
SW spectra of R-$\cri$ are calculated within QSGW with a $8\times 8 \times 8$ $\bfq$-point mesh and a $12 \times 12 \times 12$ $\bfq$-point mesh.
}
\label{fig:x2_dftu}
\end{figure}

\begin{figure}
\begin{tabular}{c}
\includegraphics[width=0.75\linewidth,clip]{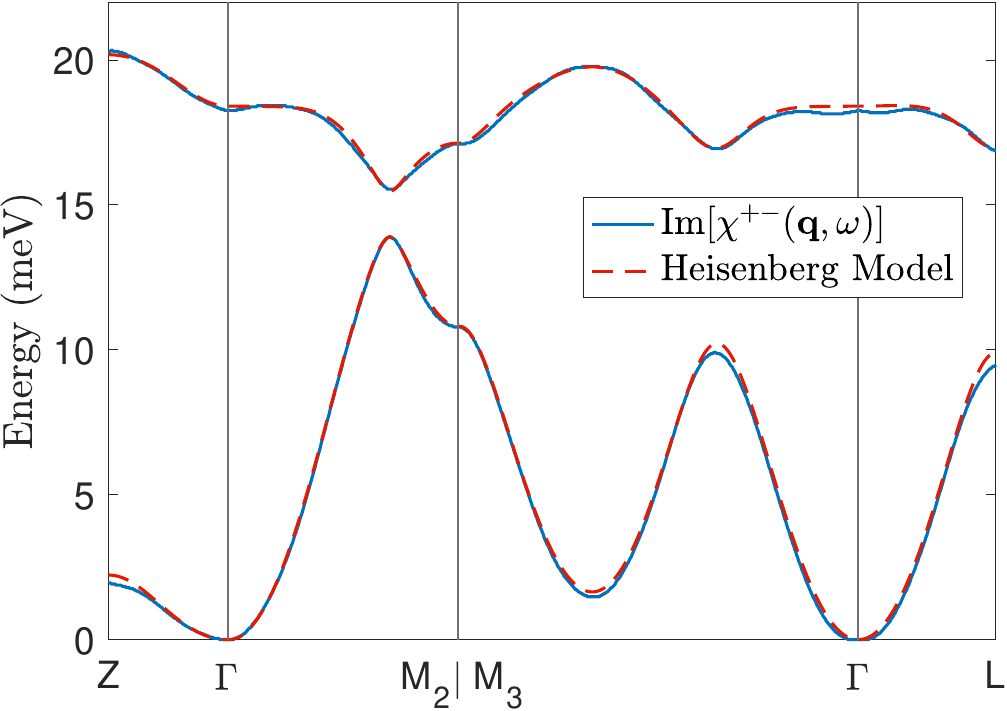} 
\end{tabular}%
\caption{Comparison of SW spectra of $\cri$ calculated from the peak of ${\chi^{+-} \left( {\bf q}, \omega \right)}$ and the Heisenberg model.
 Calculations were carried out within QSGW@DFT+$U$ with $U$=1.36 eV. 
 The SW can be reasonably described using the Heisenberg model. 
 Exchange couplings up to \SI{12}{\AA} are included in the Heisenberg model.
}
\label{fig:x2_vs_heisenberg}
\end{figure}

\begin{figure}
\begin{tabular}{c}
\includegraphics[width=0.75\linewidth,clip]{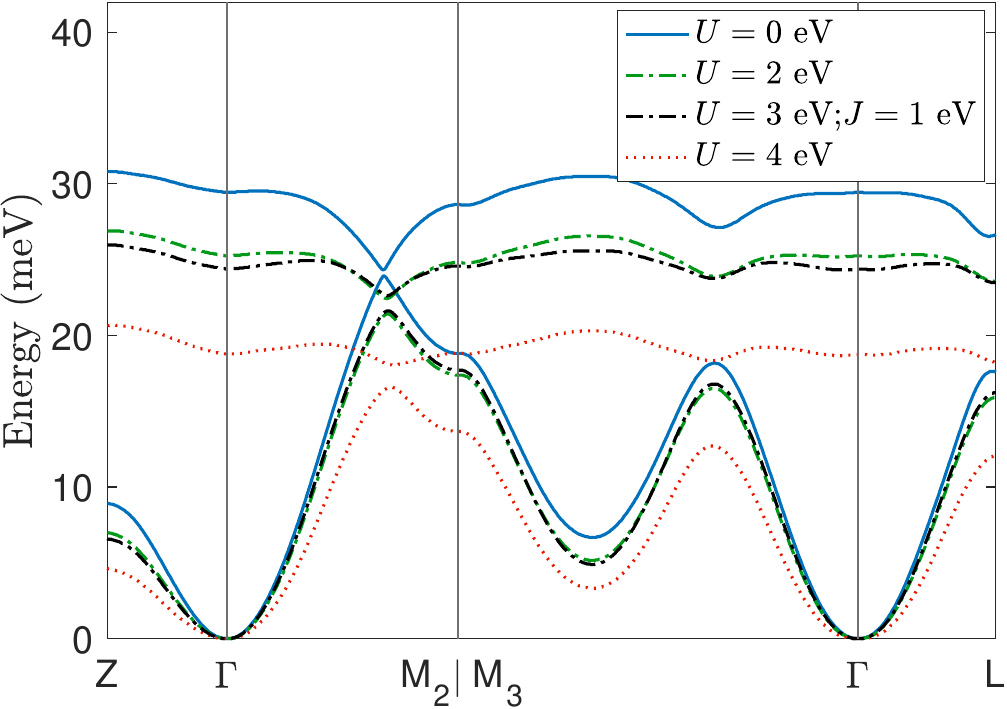} 
\end{tabular}%
\caption{The dependence of SW energy of R-$\cri$ on the $U$ parameter within DFT+$U$.
  The fully-localized-limit (FLL) double counting scheme was employed.
SW spectra are also calculated using $U=$ 0, 2, and 4 eV, and $U=\SI{3}{\eV}$ \& $J_\text{H}=\SI{1}{\eV}$. 
With $U=\SI{4}{\eV}$, the optical magnon has similar energy as in experiments but is less dispersive.
The SW energy of the acoustic mode, especially along the $\Gamma$--$Z$ ($k_z$) direction is overestimated, in comparison to experiments~\cite{chen2018prx}.
With $U=$\SI{3}{\eV} (shown in the main text), applying $J_\text{H}=\SI{1}{\eV}$ decreases the on-site Cr moment and increases the SW energy.
The resulted SW spectra are similar to those obtained with $U=\SI{2}{\eV}$ and $J_\text{H}=\SI{0}{\eV}$.
}
\label{fig:x2_dftu}
\end{figure}

\begin{figure}
\begin{tabular}{c}
  \includegraphics[width=0.75\linewidth,clip]{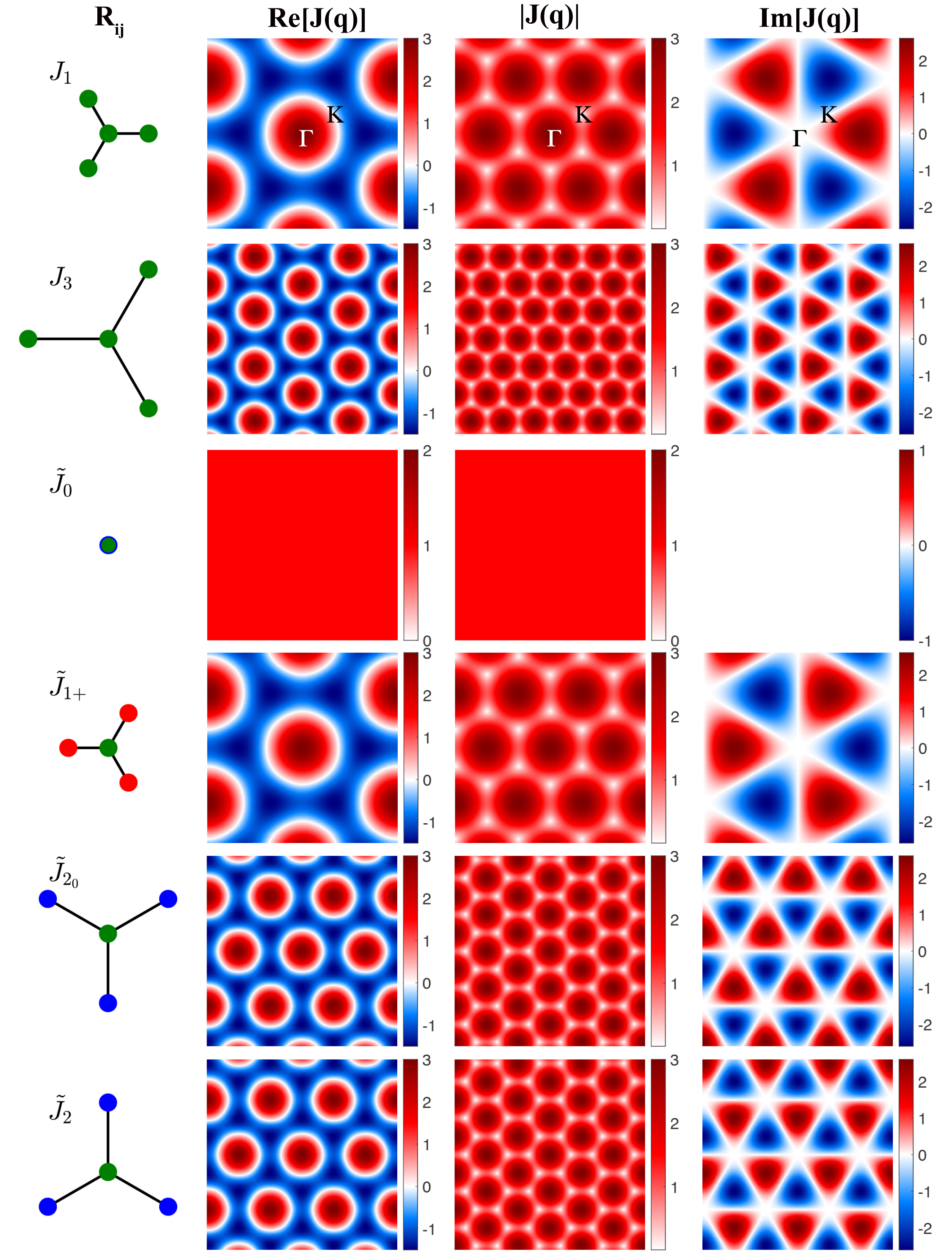} 
\end{tabular}%
\caption{Inter-sublattice exchange couplings ${J_{ij}}$ and corresponding Fourier Transform ${J(q)}$.
  The first column shows the top view of the connecting vectors $R_{ij}$ of the real-space exchange ${J_{ij}}$; Cr atoms at the upper, middle, and lower layer are in red, green, and blue, respectively. 
Among the interlayer couplings,  $\tilde{J}_0$ is the only vertical coupling, while all others have in-plane components in their connecting vectors $R_{ij}$.
In comparison to $J_1$, $J_3$, and $\tilde{J}_{1+}$, the in-plane components of connecting vectors of $\tilde{J}_{2}$ and $\tilde{J}_{2_0}$ are rotated by $\pi/6$.
The other three columns show the real part, imaginary part, and norm of the corresponding $J(\bfq)$ with $\bfq$ in the $k_x$-$k_y$ plane. 
$J_1(\bfq)$, $J_3(\bfq)$, and $J_{1+}(\bfq)$ are real functions along the $\Gamma$--$K$ path and vanishes at the point $K$.
$\tilde{J}_0(\bfq)$ is a real constant in the $k_x$-$k_y$ plane.
$J_2(\bfq)$ and $J_{2_0}(\bfq)$ are complex numbers along the $\Gamma$--$K$ path.
}
\label{fig:xtal2}
\end{figure}.

\clearpage
\section{Supplementary Tables}

\begin{table}[hbt]
  \caption{{Atomic spin magnetic $m_i$ and Curie temperatures $T_\text{C}$ in R-$\cri$ calculated using various methods.} $T_\text{C}$ are calculated using mean-field theory (MFT) and random phase approximation (RPA).
The calculated $T_\text{C}^\text{RPA}=\SI{71.8}{\K}$ using exchange parameters obtained from QSGW+$U$ compares well with previously reported~\cite{mcguire2015cm} experimental value of $T_\text{C}^\text{Exp}=\SI{61}{\K}$.}
\label{tbl:mi}
\bgroup
\def\arraystretch{1.1}
\begin{tabular*}{\linewidth}{l@{\extracolsep{\fill}}cccccccc}
\hline\hline
  {Method}   & $U$ & $J_\text{H}$ &  \multicolumn{2}{c}{$m_{i}$ ($\mu_\text{B}$/atom)}  &   & \multicolumn{2}{c}{$T_\text{C}$ (K)} \\
  \cline{2-2} \cline{3-3} \cline{4-5} \cline{7-8} 
           & (eV) & (eV) & Cr   & I                                              &   & MFT & RPA \\
\hline
DFT      & 0       &   & 3.02 & -0.07  & & 126.1  & 100.0  \\
         & 1       &   & 3.11 & -0.09  & & 125.4  &  99.9  \\
         & 2       &   & 3.20 & -0.11  & & 117.3  &  94.6  \\
         & 3       & 1 & 3.20 & -0.11  & & 116.0  &  94.2  \\  
         & 3       &   & 3.29 & -0.14  & & 106.4  &  87.3  \\
         & 4       &   & 3.38 & -0.16  & &  94.4  &  79.6  \\
         & 5       &   & 3.47 & -0.18  & &  83.1  &  72.7  \\
QSGW     &         &   & 3.10 & -0.09  & &  93.6  &  77.5  \\
         & 1.36    &   & 3.21 & -0.11  & &  82.8  &  71.8  \\
\hline\hline
\end{tabular*}
\egroup
\end{table}

\begin{table}
\caption{{Pairwise exchange parameters in R-$\cri$.}
$J_{ij}$ for the Heisenberg Hamiltonian $H=-\sum_{i\ne j} J_{ij}\,\hat{\bf e}_i \cdot \hat{\bf e}_j$, and $\hat{\bf e}_i$ is the unit vector of the local spin moment at site $i$.
The connecting vectors $\hat{\bf R}_{ij}$ are in the unit of lattice constant $a=\SI{3.965}{\AA}$.
The distance between two sites---$R_{ij}$---is in the unit of \AA\ and $a$.
Intralayer and interlayer couplings are labeled as $J_{i}$ and $\tilde{J}_{i}$, respectively.
The numbers of corresponding exchange pairs are also shown.
Symbol * denotes intra-sublattice couplings, which do not contribute the magnon gap between acoustic and optical modes.
$U=\SI{3}{\eV}$ and $U=$\SI{1.36}{\eV} are used in DFT+$U$ with QSGW+$U$ calculations, respectively.
}
\sepfootnotecontent {a}{Symbol * denotes $J_{ij}(R)$ with sites $i=j$ in the primitive cell.}
\sepfootnotecontent {b}{Fitting INS with a $J_1$-$J_2$-$J_3$-$\tilde{J}_0$ model, $J_{c1}$ is the effective $J_c$ between layers, which includes $J_{c2}$ and $J_{c3 }$.}
\label{tbl:jij}%
\bgroup
\def\arraystretch{1.1}
\small
\begin{tabular*}{\linewidth}{llc@{\extracolsep{\fill}}ccrrrrrrrrrr}
  \hline  \hline
  \\[-1em]
  \multicolumn{2}{c}{R-$\cri$} & & \multicolumn{2}{c}{$R_{ij}$}  &  & \multicolumn{3}{c}{$\hat{\bf R}_{ij}$}   &  & \multicolumn{4}{c}{$J_{ij}$(meV)}\\
  \\[-1em]
  \cline{1-2} \cline{4-5}  \cline{7-9}   \cline{11-14}
  \\[-1.1em]
  Lbl.  & No. & & \AA & $a$ &  & $x$ & $y$ & $z$ &  & DFT & DFT+$U$ & $GW$ & $GW$+$U$ \\
  \\[-1.1em]
  \hline
  \\[-1.1em]
      $J_{1}$           & 3  & &  3.965  & 1.000 & &    1     &  0     &  0.003 &  & 3.29  &  2.67 &  2.84 &  2.48  \\
                        &    & &         &       & &   -0.5   &  0.866 &  0.003 &  &       &       &       &        \\
                        &    & &         &       & &   -0.5   & -0.866 &  0.003 &  &       &       &       &        \\
      $J_2$             & 6* & &  6.867  & 1.732 & &    0     &  1.732 &  0     &  &  0.57 &  0.61 &  0.43 &  0.40  \\
                        &    & &         &       & &    0     & -1.732 &  0     &  &       &       &       &        \\
                        &    & &         &       & &   -1.5   & -0.866 &  0     &  &       &       &       &        \\
                        &    & &         &       & &    1.5   &  0.866 &  0     &  &       &       &       &        \\
                        &    & &         &       & &   -1.5   &  0.866 &  0     &  &       &       &       &        \\
                        &    & &         &       & &    1.5   & -0.866 &  0     &  &       &       &       &        \\
      ${J}_{3}$         & 3  & &  7.929  & 2.000 & &   -2     &  0     &  0.003 &  & -0.07 &  0.02 & -0.06 & -0.07  \\
                        &    & &         &       & &    1     & -1.732 &  0.003 &  &       &       &       &        \\
                        &    & &         &       & &    1     &  1.732 &  0.003 &  &       &       &       &        \\  
  \\[-1.1em]  
  \hline
  \\[-1.1em]
      $\tilde{J}_{0}$   & 1  & &  6.589  & 1.662 & &    0     &  0     & -1.662 &  &  0.19 &  0.25 &  0.14 &  0.14   \\  
      $\tilde{J}_{1-}$  & 6* & &  7.701  & 1.942 & &    1     &  0     &  1.665 &  &  0.31 &  0.25 &  0.16 &  0.13       \\
                        &    & &         &       & &   -0.5   &  0.866 &  1.665 &  &       &       &       &             \\
                        &    & &         &       & &   -0.5   & -0.866 &  1.665 &  &       &       &       &             \\  
                        &    & &         &       & &   -1     &  0     & -1.665 &  &       &       &       &             \\
                        &    & &         &       & &    0.5   & -0.866 & -1.665 &  &       &       &       &             \\
                        &    & &         &       & &    0.5   &  0.866 & -1.665 &  &       &       &       &             \\
      $\tilde{J}_{1+}$  & 3  & &  7.713  & 1.945 & &   -1     &  0     &  1.669 &  &  0.47 &  0.42 &  0.26 &  0.23       \\
                        &    & &         &       & &    0.5   & -0.866 &  1.669 &  &       &       &       &             \\
                        &    & &         &       & &    0.5   &  0.866 &  1.669 &  &       &       &       &             \\
      $\tilde{J}_{2_0}$ & 3  & &  9.517  & 2.400 & &    0     & -1.732 & -1.662 &  &  0.00 &  0.00 &  0.00 &  0.00       \\
                        &    & &         &       & &    1.5   &  0.866 & -1.662 &  &       &       &       &             \\
                        &    & &         &       & &   -1.5   &  0.866 & -1.662 &  &       &       &       &             \\
      $\tilde{J}_{2}$   & 3  & &  9.517  & 2.400 & &    0     &  1.732 & -1.662 &  & -0.04 & -0.19 & -0.18 & -0.25       \\    
                        &    & &         &       & &   -1.5   & -0.866 & -1.662 &  &       &       &       &             \\
                        &    & &         &       & &    1.5   & -0.866 & -1.662 &  &       &       &       &             \\
  \hline\hline
\end{tabular*}
\egroup
\end{table}

\clearpage
\section{Supplementary Methods}

\paragraph*{\textbf{Linear response method.}}
Starting from an $\abinitio$ single-particle Hamiltonian $H$, such as the LDA Hamiltonian, the non-interacting transverse spin susceptibility $\xqwtb$ can be calculated using the eigenvalues and eigenfunctions of $H$ within linear response theory~\cite{karlsson2000prb,kotani2008jpcm,ke2011prbr}.
\begin{eqnarray}
\chi_0^{+-}(\bfr,\bfr',\bfq, \omega) &=&
 \sum^{\rm  occ}_{\bfk n \ispone} \sum^{\rm unocc}_{\bfk' n'\isptwo}
\frac{
\Psi_{\bfk n\ispone}^*(\bfr)      \Psi_{\bfk' n'\isptwo}(\bfr)
\Psi_{\bfk' n'\isptwo}^*(\bfr') \Psi_{\bfk n\ispone}(\bfr')
}{\omega-(\epsilon_{\bfk' n'\isptwo}-\epsilon_{\bfk n\ispone})+i \delta} \nonumber\\ 
&+& \sum^{\rm  unocc}_{\bfk n \ispone} \sum^{\rm occ}_{\bfk' n'\isptwo}
\frac{
\Psi_{\bfk n\ispone}^*(\bfr)      \Psi_{\bfk' n'\isptwo}(\bfr)
\Psi_{\bfk' n'\isptwo}^*(\bfr') \Psi_{\bfk n\ispone}(\bfr')
}{-\omega-(\epsilon_{\bfk n\ispone}-\epsilon_{\bfk' n'\isptwo})+i \delta},
\label{generalchi01q}
\end{eqnarray}
where $\bfk' = \bfq+ \bfk$. Here we assume the response is stationary and system posseses translational symmetry.
Within RPA, the full susceptibilities are calculated using $\chi=\chi_0 + \chi_0 I \chi$.
Within time-dependent LDA, the exchange correlation kernel can be explicitly calculated using $I_{\rm xc}^{+-}={B_{\rm xc}(\bfr)}/{m}(\bfr)$ for a collinear spin structure.
In general, we represent two-particle quantities $\chi^{0}$, $\chi$ and  $I$ using a mixed basis~\cite{kotani2007prb} $\{B_{i}(\bfr)\}$, which consists of the product basis~\cite{aryasetiawan1994prbA,kotani2007prb} within the augmentation spheres and interstitial plane waves. 
The product basis itself is constructed by the products of wavefunctions and their energy derivatives within the augmentation spheres.
Since magnetic moment and response are nearly completely confined within the magnetic Cr sites, we further project $\chi( \mathbf{q},\omega )$  onto  the functions $\{m_i(\bfr)/||m_i||\}$ representing normalized local spin densities on each magnetic Cr site, which gives a matrix $\chi _{ij}( \mathbf{q},\omega )$ in  magnetic basis site indices~\cite{kotani2008jpcm}.
This projection corresponds to the rigid spin approximation (RSA), which is a modest approximation for $\cri$.

The sum rule~\cite{kotani2008jpcm} used can be written as 
\begin{equation}
  \frac{M(\bfra)}{\omega} =  \int_\Omega \ud \bfrb \ \chi^{+-} (\bfra,\bfrb,\bfq=0,\omega),
\label{eqsumm}
\end{equation}
where $\Omega$ denotes the unit-cell volume.
Within RSA, \req{eqsumm} reduce to two equations for each $\omega$.
We further assume that kernel $I(\omega)$ is site-diagonal and $\bfq$-independent, allowing it to be determined solely by \req{eqsumm}.
Note that \req{eqsumm} ensures the Goldstone magnon mode at $\bfq=0$ and $\omega=0$.

To obtain a high-resolution SW spectra along the high symmetry path, we need to calculate $\chi^{+-}({\bf q},\omega)$ with a dense set of $\bfq$ points along the path.
We first calculate  $\chi_0^{+-}({\bf q},\omega)$ on a $8 \times 8 \times 8$ $\bfq$ mesh in the first BZ.
Using a fast Fourier transform, we obtain $\chi_0^{+-}({\bf R},\omega)$ on a real-space mesh.
Then we calculate the $\chi_0^{+-}({\bf q},\omega)$ along the high symmetry path by Fourier transforming $\chi_0^{+-}({\bf R},\omega)$.
A denser $12 \times 12 \times 12$ $\bfq$(${\bf R}$) mesh does not improve the results (See Fig.~S1).

Future investigation of spin susceptibilities going beyond RSA may be valuable. 
For example, a complete basis set can be used to take into account of the deformation of the magnetic moment in spin excitation and allow 
a more comprehensive description of magnetic interaction beyond the Heisenberg model.

\paragraph*{\textbf{Single-particle Hamiltonian.}}
To include electron correlation effects beyond DFT, we use DFT$+U$ and $GW$ methods to describe the electronic structure and prepare the one-particle Hamiltonian.
DFT+$U$ provides the simplest form to include the on-site non-local correlation effects, while the $GW$ method explicitly include the electron-electron interaction by calculating the dynamically-screened Coulomb interaction $W$.
$GW$ methods have been applied on top of DFT or DFT+$U$~\cite{jiang2012prb} for various systems.
Different flavors of $GW$-based methods have been developed.
Here, we use the recently developed quasiparticle self-consistent $GW$ ($\qsgw$) method~\cite{van-schilfgaarde2006prl}, which is 
a self-consistent perturbation method based on the $GW$ approximation (GWA)~\cite{hedin1969book}. 
Applying GWA on the top of the Kohn-Sham Hamiltonian $H_0$ of DFT, the self-energy is calculated as $\Sigma({\bf r},{\bf r}',\omega)=\Sigma(1,2) =iG_0(1,2)W(1+,2)$.
Here $G_0$ is the Green's function of $H_0$, and $W$ is the dynamically-screened Coulomb interaction calculated using $G_0$ within random phase approximation.
Then self-consistent calculation is performed by replacing the exchange-correlation term $V^{\rm xc}_{\rm DFT} $ in DFT  with $ V^{\rm xc}_{\rm QSGW}(\bf r,\bf r')$, whose matrix elements are given as
\begin{equation}
V^{\rm xc}_{\rm QSGW} = \frac{1}{2}\sum_{ij} |\psi_i\rangle 
       \left\{ {{\rm Re}[\Sigma(\epsilon_i)]_{ij}+{\rm Re}[\Sigma(\epsilon_j)]_{ij}} \right\}
       \langle\psi_j|,  
\label{eq:vxc}
\end{equation}
where $\epsilon_i$ and $|\psi_i\rangle$ are the eigenvalues and
eigenfunctions of $H_0$, respectively, and
$\Sigma_{ij}(\omega) = \langle \psi_i|\Sigma(\omega)| \psi_j \rangle =
\int d^3r \int d^3r' \psi_i^*({\bf r}) \Sigma({{\bf r}},{{\bf r}'},\omega)
\psi_j({\bf r}')$.
${\rm Re}[\Sigma(\omega)]$ is the Hermitian part of the
self-energy~\cite{kotani2007prb}.

DFT+$U$ calculations are performed within the generalized gradient approximation (GGA) using the exchange-correlation functional of Perdew, Burke, and Ernzerhof (PBE)~\cite{perdew1996prl}.
The fully-localized-limit (FLL)~\cite{liechtenstein1995prb} double counting scheme was employed.
Multiple Hubbard $U$ values (0--5 eV) are used to simulate the additional Cr $d$-orbital on-site correlations not accounted for in DFT calculations.
To understand the effect of Hund's Exchange $J_\text{H}$, we also carry out calculations with $U=\SI{3}{\eV}$ and $U_\text{H}=\SI{1}{\eV}$.
As shown in \rtbl{tbl:mi} and \rfig{fig:x2_dftu}, the calculated on-site magnetic moments, Curie temperature, and SW spectra are very similar to those obtained using $U=\SI{2}{\eV}$ and $U_\text{H}=\SI{0}{\eV}$.
Thus, the Hund's Exchange lowers the on-site moments of Cr and I atoms and increases the SW energy and Curie temperature.

All DFT, DFT+$U$ and {\qsgw} calculations were performed using a full-potential generalization~\cite{methfessel2000coll-c3fl,pashov2020cpc} of the standard linear muffin-tin orbital (LMTO) basis set~\cite{andersen1975prb}.
This scheme employs generalized Hankel functions as the envelope functions.
Experimental crystal structures~\cite{mcguire2015cm} are used and spin-orbit coupling (SOC) is not included in all calculations.
Further details of QSGW implementation~\cite{van-schilfgaarde2006prl, kotani2007prb, kotani2014jpsj} and applications on $\cri$~\cite{lee2020prb} can be found elsewhere.

\section{Supplementary Discussions}
\paragraph*{\textbf{The effects of SOC.}}
Both Dzyaloshinskii-Moriya interaction (DMI) and magnetocrystalline anisotropy (MA) originate from SOC.
The effects of MA and possible DMI on spinwave spectra in these systems have been widely investigated in the linear spinwave theory.
Including MA, either in a single-ion or two-ion form (anisotropic exchange), introduces a gap in the spinwave excitation at the $\Gamma$ point.
Using INS, Chen $\etal$~\cite{chen2020prb} have demonstrated that MA plays a dominant role in stabilizing the ferromagnetic ordering in $\cri$ and found a MA-induced gap of \SI{0.37}{\meV} at $T=\SI{3}{\K}$.
On the other hand, including the second-nearest-neighbor DMI in the Heisenberg model Hamiltonian opens a global magnon gap through the whole BZ.

\input{supplementary.bbl}

\end{document}

%% file: supplementary.bbl
%